\documentclass[a4paper,10pt,twoside]{article}
\usepackage[INRIA]{lipinria}
\usepackage[english]{babel}
\usepackage[latin1]{inputenc}
\usepackage[T1]{fontenc}
\usepackage{textcomp, amsthm, amsmath,amsfonts, amssymb, latexsym, amscd}
\usepackage{figlatex}
\usepackage{rotating}
\usepackage{url}
\newcommand{\Ss}[1]{\ensuremath{Comm^{start}_{#1}}}
\newcommand{\Se}[1]{\ensuremath{Comm^{end}_{#1}}}
\newcommand{\Cs}[1]{\ensuremath{Comp^{start}_{#1}}}
\newcommand{\Ce}[1]{\ensuremath{Comp^{end}_{#1}}}
\newcommand{\fr}[3]{\ensuremath{\gamma^{#2}_{#1}(#3)}}

\newcommand{\NSs}[1]{\ensuremath{Comm'\,^{start}_{#1}}}
\newcommand{\NSe}[1]{\ensuremath{Comm'\,^{end}_{#1}}}
\newcommand{\NCs}[1]{\ensuremath{Comp'\,^{start}_{#1}}}
\newcommand{\NCe}[1]{\ensuremath{Comp'\,^{end}_{#1}}}

\newtheorem{theorem}{Theorem}

\begin{document}

\RRInumber{2007-07}

\bibliographystyle{plain}
\RRItitle{Comments on ``Design and performance evaluation of load distribution strategies for multiple loads on heterogeneous linear daisy chain networks''}
\RRItitre{Commentaires sur «~Design and performance evaluation of load distribution strategies for multiple loads on heterogeneous linear daisy chain networks~»}
\RRIthead{Comments on ``Design performance evaluation of load distribution strategies...''}

\RRIdate{February 2007}

\RRIauthor{Matthieu Gallet \and Yves Robert \and Frédéric Vivien}
\RRIahead{M. Gallet \and Y. Robert \and F. Vivien}

\RRIkeywords{scheduling, heterogeneous processors, divisible loads, single-installment, multiple-installments.}
\RRImotscles{ordonnancement, ressources hétérogènes, tâches divisibles, tournées.}

\RRIabstract{Min, Veeravalli, and Barlas
  proposed~\cite{WongVe04,WongVeBa05} strategies to minimize the
  overall execution time of one or several divisible loads on a
  heterogeneous linear network, using one or more installments. We
  show on a very simple example that the approach proposed
  in~\cite{WongVeBa05} does not always produce a solution and that,
  when it does, the solution is often suboptimal.  We also show how to
  find an optimal scheduling for any instance, once the number of
  installments per load is given. Finally, we formally prove that
  under a linear cost model, as in~\cite{WongVe04,WongVeBa05}, an
  optimal schedule has an infinite number of installments. Such a cost
  model can therefore not be used to design practical
  multi-installment strategies.}

\RRIresume{Min, Veeravalli, and Barlas ont
  proposé~\cite{WongVe04,WongVeBa05} des stratégies pour minimiser le
  temps d'exécution d'une ou de plusieurs tâches divisibles sur un
  réseau linéaire de processeurs hétérogènes, en distribuant le
  travail en une ou plusieurs tournées. Sur un exemple très simple
  nous montrons que l'approche proposée dans~\cite{WongVeBa05} ne
  produit pas toujours une solution et que, quand elle le fait, la
  solution est souvent sous-optimale. Nous montrons également comment
  trouver un ordonnancement optimal pour toute instance, quand le
  nombre de tournées par tâches est spécifié. Finalement, nous
  montrons formellement que lorsque les fonctions de coûts sont
  linéaires, comme c'est le cas dans~\cite{WongVe04,WongVeBa05}, un
  ordonnancement optimal a un nombre infini de tournées. Un tel modèle
  de coût ne peut donc pas être utilisé pour définir des stratégies en
  multi-tournées utilisables en pratique.}

\RRItheme{\THNum}
\RRIprojet{GRAAL}

\RRImaketitle

\section{Introduction}
\label{sec.intro}

Min, Veeravalli and Barlas proposed~\cite{WongVe04,WongVeBa05}
strategies to minimize the overall execution time of one or several
divisible loads on a heterogeneous linear network. Initially, the
authors targeted single-installment strategies, that is strategies
under which a processor receives in a single communication all its
share of a given load. When they were not able to design
single-installment strategies, they proposed multi-installment ones.

In this research note, we first show on a very simple example that the
approach proposed in~\cite{WongVeBa05} does not always produce a
solution and that, when it does, the solution is often suboptimal.
The fundamental flaw of the approach of~\cite{WongVeBa05} is that the
authors are optimizing the scheduling load by load, instead of
attempting a global optimization. The load by load approach is
suboptimal and overconstrains the problem.

On the contrary, we show how to find an optimal scheduling for any
instance, once the number of installments per load is given. In
particular, our approach always find the optimal solution in the
single-installment case. Finally, we formally prove that under a
linear cost model for communication and communication, as
in~\cite{WongVe04,WongVeBa05}, an optimal schedule has an infinite
number of installments. Such a cost model can therefore not be used to
design practical multi-installment strategies.

Please refer to the papers~\cite{WongVe04,WongVeBa05} for a detailed
introduction to the optimization problem under study. We briefly
recall the framework in Section~\ref{sec.frame}, and we deal with an
illustrative example in Section~\ref{sec.example}. Then we directly
proceed to the design of our solution (Section~\ref{sec.new}), we
discuss its possible extensions and the linear cost model
(Section~\ref{sec:ext}), before concluding
(Section~\ref{sec:conclusion}).

\section{Problem and Notations}
\label{sec.frame}

\begin{figure}
\begin{center}
	{\includegraphics[width=0.6\linewidth]{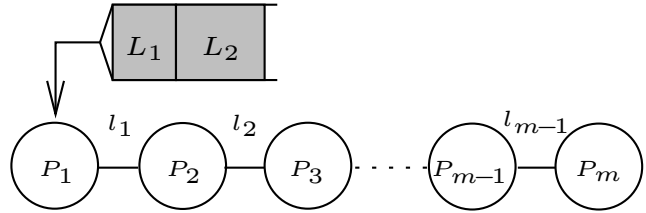}}
        \caption{Linear network, with $m$ processors and $m-1$ links.}
        \label{Fig:platform}
\end{center}
\end{figure}

We summarize here the framework of~\cite{WongVe04,WongVeBa05}.  The
target architecture is a linear chain of $m$ processors ($P_{1}, P_2,
\ldots, P_{m}$).  Processor $P_{i}$ is connected to processor
$P_{i+1}$ by the communication link $l_{i}$ (see
Figure~\ref{Fig:platform}).  The target application is composed of $N$
loads, which are \emph{divisible}, which means that each load can be
split into an arbitrary number of chunks of any size, and these chunks
can be processed independently.  All the loads are initially available
on processor $P_{1}$, which processes a fraction of them and delegates
(sends) the remaining fraction to $P_2$.  In turn, $P_2$ executes part
of the load that it receives from $P_1$ and sends the rest to $P_3$,
and so on along the processor chain. Communications can be overlapped
with (independent) computations, but a given processor can be active
in at most a single communication at any time-step: sends and receives
are serialized (this is the full \emph{one-port} model).

Since the last processor $P_m$ cannot start computing before having
received its first message, it is useful for $P_1$ to distribute the
loads in several installments: the idle time of remote processors in
the chain will be reduced due to the fact that communications are
smaller in the first steps of the overall execution.

We deal with the general case in which the $n$th load is distributed
in $Q_{n}$ installments of different sizes. For the $j$th installment
of load $n$, processor $P_i$ takes a fraction \fr{j}{n}{i}, and sends
the remaining part to the next processor while processing its own
fraction.

In the framework of~\cite{WongVe04,WongVeBa05}, loads have different
characteristics.  Every load $n$ (with $1 \leq n \leq N$) is defined
by a volume of data $V_{comm}(n)$ and a quantity of computation
$V_{comp}(n)$. Moreover, processors and links are not identical
either. We let $w_{i}$ be the time taken by $P_{i}$ to compute a unit
load ($1\leq i\leq m$), and $z_{i}$ be the time taken by $P_{i}$ to
send a unit load to $P_{i+1}$ (over link $l_i$, $1\leq i \leq m-1$).
Note that we assume a linear model for computations and
communications, as in the original articles, and as is often the case
in divisible load
literature~\cite{Robertazzi-Computer2003,cluster-special}.

For the $j$th installment of the $n$th load, let $\Ss{i,n,j}$ denote
the starting time of the communication between $P_{i}$ and $P_{i+1}$,
and let $\Se{i,n,j}$ denote its completion time; similarly,
$\Cs{i,n,j}$ denotes the start time of the computation on $P_{i}$ for
this installment, and $\Ce{i,n,j}$ denotes its completion time.  The
objective function is to minimize the \emph{makespan}, i.e., the time
at which all loads are computed.  For the sake of convenience, all
notations are summarized in Table~\ref{notations}.

\begin{table*}
\begin{center}
{\small
\begin{tabular}{|ll|}
\hline
$m$ & Number of processors in the system.\\
\hline
$P_{i}$ & Processor $i$, where $i = 1, \ldots, m$.\\
\hline
$w_{i}$ & Time taken by processor $P_{i}$ to compute a unit load.\\
\hline
$z_{i}$ & Time taken by $P_{i}$ to transmit a unit load to $P_{i+1}$.\\
\hline
$\tau_{i}$ & Availability date of $P_{i}$ (time at which it becomes available for processing the loads).\\
\hline
$N$ & Total number of loads to process in the system.\\
\hline
$Q_{n}$ & Total number of installments for $n$th load.\\
\hline
$V_{comm}(n)$ & Volume of data for $n$th load.\\
\hline
$V_{comp}(n)$ & Volume of computation for $n$th load.\\
\hline
$\fr{i}{j}{n}$ & Fraction of $n$th load computed on processor $P_{i}$ during the $j$th installment.\\
\hline
$\Ss{i,n,j}$ & Start time of communication from processor $P_{i}$ to processor $P_{i+1}$\\
& for $j$th installment of $n$th load.\\
\hline
$\Se{i,n,j}$ & End time of communication from processor $P_{i}$ to processor $P_{i+1}$\\
& for $j$th installment of $n$th load.\\
\hline
$\Cs{i,n,j}$ & Start time of computation on processor $P_{i}$\\
& for $j$th installment of $n$th load.\\
\hline
$\Ce{i,n,j}$ & End time of computation on processor $P_{i}$\\
& for $j$th installment of $n$th load.\\
\hline
\end{tabular}}
\caption{Summary of notations.\label{notations}}
\end{center}
\end{table*}

\section{An illustrative example}
\label{sec.example}

\subsection{Presentation}
\label{sec.ours}

To show the limitations of~\cite{WongVe04,WongVeBa05}, we deal with a simple illustrative example.
We use $2$ identical processors $P_{1}$ and $P_{2}$ with $w_{1} = w_{2} = \lambda$, and $z(1) = 1$.
We consider $N= 2$ identical divisible loads to process, with $V_{comm}(1)=V_{comm}(2)=1$ and $V_{comp}(1)=V_{comp}(2)=1$. Note that when $\lambda$ is large, communications become negligible and each
processor is expected to process around half of both loads. But when $\lambda$ is close to $0$,
communications are very important, and the solution is not obvious. To ease the reading, we only
give a short (intuitive) description of the schedules, and provide their different makespans without justification
(we refer the reader to Appendix~\ref{annexe} for all proofs).

\begin{figure}
\begin{center}
	{\includegraphics[width=0.6\linewidth]{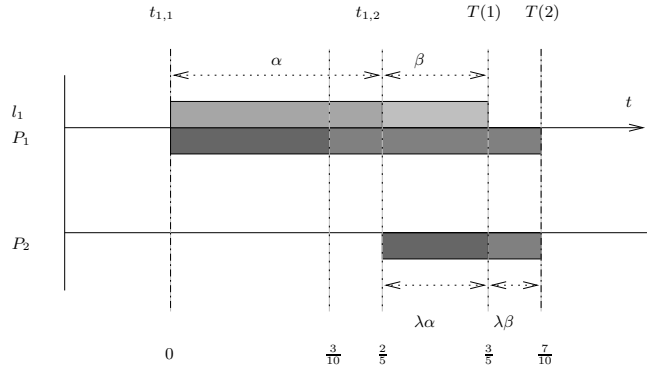}}
        \caption{The example schedule, with $\lambda=\frac{1}{2}$, $\alpha$ is $\fr{2}{1}{1}$ and $\beta$ is $\fr{2}{1}{2}$.\label{Fig:example}}
\end{center}
\end{figure}

We first consider a simple schedule which uses a single installment for each load,
as illustrated in Figure \ref{Fig:example}.
Processor $P_{1}$ computes a fraction $\fr{1}{1}{1} =	\frac{2\lambda^{2}+1}{2\lambda^{2}+2\lambda+1}$ of the first load, and a fraction $\fr{1}{1}{2}	= \frac{2\lambda+1}{2\lambda^{2}+2\lambda+1}$ of the second load.
Then the second processor computes a fraction $\fr{2}{1}{1} = \frac{2\lambda}{2\lambda^{2}+2\lambda+1}$ of the first load, and a fraction $\fr{2}{1}{2}	= \frac{2\lambda^{2}}{2\lambda^{2}+2\lambda+1}$ of the second load.
The makespan achieved by this schedule is equal to $\mathrm{makespan}_{1}=\frac{2\lambda\left(\lambda^{2}+\lambda+1\right)}{2\lambda^{2}+2\lambda+1}$.

\subsection{Solution of~\cite{WongVeBa05}, one-installment}
\label{oneinstallment}

In the solution of~\cite{WongVeBa05}, $P_{1}$ and $P_{2}$ have to
simultaneously complete the processing of their share of the first
load. The same holds true for the second load. We are in the
one-installment case when $P_1$ is fast enough to send the second load
to $P_{2}$ while it is computing the first load. This condition writes
$\lambda\geq \frac{\sqrt{3}+1}{2}\approx 1.366$.

In the solution of~\cite{WongVeBa05}, $P_{1}$ processes a fraction
$\fr{1}{1}{1}=\frac{\lambda+1}{2\lambda+1}$ of the first load, and a
fraction $\fr{1}{1}{2}=\frac{1}{2}$ of the second one. $P_{2}$
processes a fraction $\fr{2}{1}{1}=\frac{\lambda}{2\lambda+1}$ of the
first load $L_{1}$, and a fraction $\fr{2}{1}{2}=\frac{1}{2}$ of the
second one. The makespan achieved by this schedule is
$\mathrm{makespan}_{2} =
\frac{\lambda\left(4\lambda+3\right)}{2\left(2\lambda+1\right)}$.

\begin{figure}
\begin{center}
	{\includegraphics[width=0.6\linewidth]{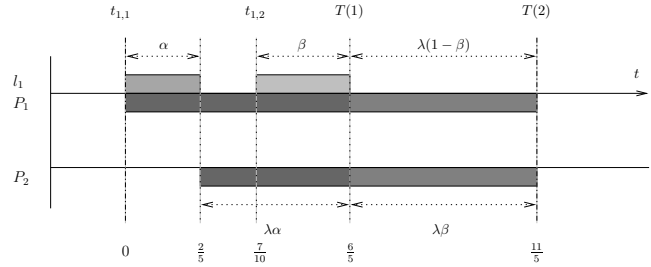}}
        \caption{The schedule of~\cite{WongVeBa05} for $\lambda=2$, with $\alpha=\fr{2}{1}{1}$
        and $\beta =\fr{2}{1}{2}$.}
        \label{Fig:example_l2}
\end{center}
\end{figure}

Comparing both makespans, we have $0\leq
\mathrm{makespan}_{2}-\mathrm{makespan}_{1}\leq \frac{1}{4}$, the
solution of~\cite{WongVeBa05} having a strictly larger makespan,
except when $\lambda = \frac{\sqrt{3}+1}{2}$. Intuitively, the
solution of~\cite{WongVeBa05} is worse than the schedule of
Section~\ref{sec.ours} because it aims at locally optimizing the
makespan for the first load, and then optimizing the makespan for the
second one, instead of directly searching for a global optimum.  A
visual representation of this case is given in
Figure~\ref{Fig:example_l2} for $\lambda=2$.

\subsection{Solution of~\cite{WongVeBa05}, multi-installment}
\label{multiinstallment}

The solution of~\cite{WongVeBa05} is a multi-installment strategy when
$\lambda < \frac{\sqrt{3}+1}{2}$, i.e., when communications tend to be
important compared to computations.  More precisely, this case happens
when $P_{1}$ does not have enough time to completely send the second
load to $P_{2}$ before the end of the computation of the first load on
both processors.

The way to proceed in~\cite{WongVeBa05} is to send the second load
using a multi-installment strategy. Let $Q$ denote the number of
installments for this second load.  We can easily compute the size of
each fraction distributed to $P_{1}$ and $P_{2}$.  Processor $P_{1}$
has to process a fraction $\fr{1}{1}{1}=\frac{\lambda+1}{2\lambda+1}$
of the first load, and fractions $\fr{1}{1}{2}, \fr{1}{2}{2},\ldots,
\fr{1}{Q}{2}$ of the second one. Processor $P_{2}$ has a fraction
$\fr{2}{1}{1}=\frac{\lambda}{2\lambda+1}$ of the first load, and
fractions $\fr{2}{1}{2}, \fr{2}{2}{2},\ldots, \fr{2}{Q}{2}$ of the
second one.  Moreover, we have the following equality for $1\leq k <
Q$:
$$\fr{1}{k}{2}=\fr{2}{k}{2}=\lambda^{k}\fr{2}{1}{1}.$$
And for $k=Q$ (the last installment), we have
$\fr{1}{Q}{2}=\fr{2}{Q}{2} \leq \lambda^{Q}\fr{2}{1}{1}$. Let
$\beta_{k}=\fr{1}{k}{2}=\fr{2}{k}{2}$. We can then establish an upper
bound on the portion of the second load distributed in $Q$
installments:
$$\sum_{k=1}^{Q}\left(2\beta_{k}\right) \leq 2\sum_{k=1}^{Q}\left(\fr{2}{1}{1}\lambda^{k}\right) =\frac{2\left(\lambda^{Q}-1\right)\lambda^{2}}{2\lambda^{2}-\lambda-1}$$
if $\lambda\neq 1$, and $Q=2$ otherwise.

We have three cases to discuss:
\begin{enumerate}
\item{$0 < \lambda < \frac{\sqrt{17}+1}{8} \approx 0.64$:}
Since $\lambda < 1$, we can write for any nonnegative integer $Q$:
$$\sum_{k=1}^{Q}\left(2\beta_{k}\right)< \sum_{k=1}^{\infty}\left(2\beta_{i}\right)=\frac{2\lambda^{2}}{(1-\lambda)(2\lambda+1)}$$
We have $\frac{2\lambda^{2}}{(1-\lambda)(2\lambda+1)} < 1$ for all
$\lambda < \frac{\sqrt{17}+1}{8}$.  So, even in the case of an
infinite number of installments, the second load will not be
completely processed.  In other words, no solution is found
in~\cite{WongVeBa05} for this case.  A visual representation of this
case is given in Figure~\ref{Fig:example_l05} with $\lambda=0.5$.
\begin{figure}
\begin{center}
	{\includegraphics[width=0.6\linewidth]{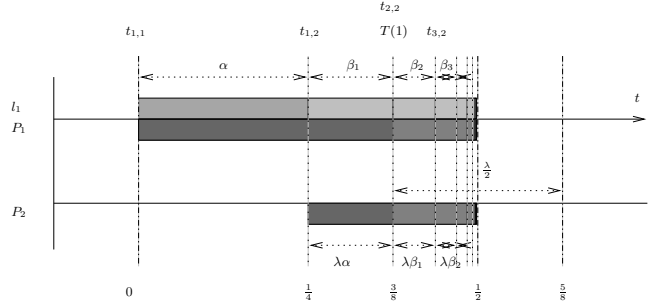}}
        \caption{The example with $\lambda=\frac{1}{2}$, $\alpha = \fr{2}{1}{1}$ and $\beta = \fr{2}{1}{2}$.} \label{Fig:example_l05}
\end{center}
\end{figure}
\item{$\lambda = \frac{\sqrt{17}+1}{8}$:}
We have $\frac{2\lambda^{2}}{(1-\lambda)(2\lambda+1)} = 1$, so an infinite number of installments is required to completely process the second load. Again, this solution is obviously not feasible.

\item{$\frac{\sqrt{17}+1}{8} < \lambda < \frac{\sqrt{3}+1}{2}$:}
In this case, the solution of~\cite{WongVeBa05} is better than any solution using a single installment per load,
but it may require a very large number of installments.
A visual representation of this case is given in Figure~\ref{Fig:example_l1} with $\lambda=1$.

\begin{figure}
\begin{center}
	{\includegraphics[width=0.6\linewidth]{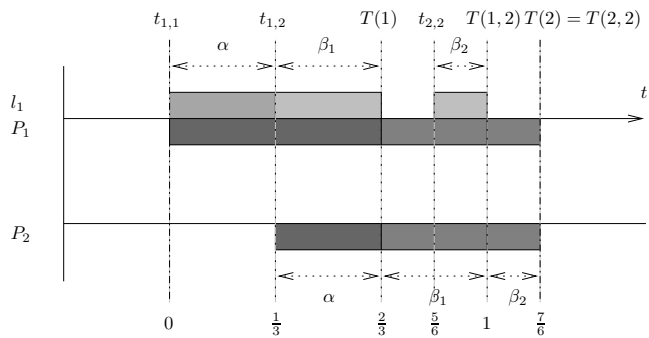}}
        \caption{The example with $\lambda=1$, $\alpha = \fr{2}{1}{1}$ and $\beta = \fr{2}{1}{2}$. \label{Fig:example_l1}}
\end{center}
\end{figure}
In this case, the number of installments is set in~\cite{WongVeBa05} as $Q = \left\lceil \frac{\ln(\frac{4\lambda^{2}-\lambda-1}{2\lambda^{2}})}{\ln(\lambda)} \right\rceil$.
To see that this choice is not optimal, consider the case $\lambda=\frac{3}{4}$.
The algorithm of~\cite{WongVeBa05} achieves a makespan equal to $\left(1-\fr{2}{1}{1}\right)\lambda+\frac{\lambda}{2}=\frac{9}{10}$. The first load is sent in one installment and the second one is sent in $3$ installments (according to the previous equation).

However, we can come up with a better schedule by splitting both loads into two installments,
and distributing them as follows:
\begin{itemize}
\item during the first round, $P_{1}$ processes $0$ unit of the first load,
\item during the second round, $P_{1}$ processes $\frac{317}{653}$ unit of the first load,
\item during the first round, $P_{2}$ processes $\frac{192}{653}$ unit of the first load,
\item during the second round, $P_{2}$ processes $\frac{144}{653}$ unit of the first load,
\item during the first round, $P_{1}$ processes $0$ unit of the second load,
\item during the second round, $P_{1}$ processes $\frac{464}{653}$ unit of the second load,
\item during the first round, $P_{2}$ processes $\frac{108}{653}$ unit of the second load,
\item during the second round, $P_{2}$ processes $\frac{81}{653}$ unit of the second load,
\end{itemize}
This scheme gives us a total makespan equal to
$\frac{781}{653}\frac{3}{4}\approx0.897$, which is (slightly) better
than $0.9$. This shows that among the schedules having a total number
of four installments, the solution of~\cite{WongVeBa05} is suboptimal.
\end{enumerate}

\subsection{Conclusion}

Despite its simplicity (two identical processors and two identical
loads), the analysis of this illustrative example clearly outlines the
limitations of the approach of~\cite{WongVeBa05}: this approach does
not always return a feasible solution and, when it does, this solution
is not always optimal. In the next section, we show how to compute an
optimal schedule when dividing each load into any prescribed number of
installments.

\section{Optimal solution}
\label{sec.new}

We now show how to compute an optimal schedule, when dividing each
load into any prescribed number of installments. Therefore, when this
number of installment is set to 1 for each load (i.e., $Q_n=1$, for
any $n$ in $[1,N]$), the following approach solves the problem
originally target by Min, Veeravalli, and Barlas.

To build our solution we use a linear programming approach. In fact,
we only have to list all the (linear) constraints that must be
fulfilled by a schedule, and write that we want to minimize the
$\mathrm{makespan}$. All these constraints are captured by the linear
program in Figure~\ref{linearprogram}. The optimality of the solution
comes from the fact that the constraints are exactly all the
constraints a schedule must fulfill, and a solution to the linear
program is obviously always feasible. This linear program simply
encodes the following constraints (where a number in brackets is the
number of the corresponding constraint on Figure~\ref{linearprogram}):
\begin{itemize}
\item $P_{i}$ cannot start a new communication to $P_{i}$ before the
  end of the corresponding communication from $P_{i-1}$ to $P_i$
  (\ref{cons1}),

\item $P_{i}$ cannot start to receive the next installment of the
  $n$th load before having finished to send the current one to
  $P_{i+1}$ (\ref{cons2}),

\item $P_{i}$ cannot start to receive the first installment of the
  next load before having finished to send the last installment of the
  current load to $P_{i+1}$ (\ref{cons3}),

\item any transfer has to begin at a nonnegative time (\ref{cons4}),

\item the duration of any transfer is equal to the product of the time
  taken to transmit a unit load (\ref{cons5}) by the volume of data to
  transfer,

\item processor $P_{i}$ cannot start to compute the $j$th installment
  of the $n$th load before having finished to receive the
  corresponding data (\ref{cons6}),

\item the duration of any computation is equal to the product of the
  time taken to compute a unit load (\ref{cons7}) by the volume of
  computations,

\item processor $P_i$ cannot start to compute the first installment of
  the next load before it has completed the computation of the last
  installment of the current load (\ref{cons8}),

\item processor $P_i$ cannot start to compute the next installment of
  a load before it has completed the computation of the current
  installment of that load (\ref{cons9}),

\item processor $P_i$ cannot start to compute the first installment of
  the first load before its availability date (\ref{cons10}),
\item every portion of a load dedicated to a processor is necessarily
  nonnegative (\ref{cons11}),
\item any load has to be completely processed (\ref{cons12}),
\item the $\mathrm{makespan}$ is no smaller than the completion time
  of the last installment of the last load on any processor
  (\ref{cons13}).
\end{itemize}

\begin{figure*}
{\small
\begin{eqnarray}
\forall i < m-1, n\leq N,j \leq Q_{n}	\quad & \Ss{i+1,n,j} 		\quad & \geq \quad \Se{i,n,j} \label{cons1}\\
\forall i < m-1,n \leq N,j < Q_{n}	\quad & \Ss{i,n,j+1}		\quad & \geq \quad \Se{i+1,n,j} \label{cons2}\\
\forall i < m-1,n < N	 		\quad & \Ss{i,n+1,1}		\quad & \geq \quad \Se{i+1,n,Q_{n}} \label{cons3}\\
\forall i \leq m-1,n\leq N,j \leq Q_{n}	\quad & \Ss{i,n,j}		\quad & \geq \quad 0\label{cons4}\\
\forall i \leq m-1,n\leq N,j \leq Q_{n}	\quad & \Se{i,n,j} 		\quad & = ~ \Ss{i,n,j} + z_{i}V_{comm}(n)\!\!\!\!\sum_{k=i+1}^{m}\!\gamma_{k}^{j}(n)\label{cons5}\\
\forall i \geq 2, n\leq N, j\leq Q_{n}	\quad & \Cs{i,n,j}		\quad & \geq \quad \Se{i,n,j} \label{cons6}\\
\forall i\leq m,n\leq N,j\leq Q_{n}	\quad & \Ce{i,n,j}		\quad & = \quad \Cs{i,n,j} + w_{i}\gamma_{i}^{j}(n)V_{calc}(n)\label{cons7}\\
\forall i\leq m,n < N			\quad & \Cs{i,n+1,1}		\quad & \geq \quad \Ce{i,n,Q_{n}} \label{cons8}\\
\forall i\leq m,n \leq N,j < Q_{n}	\quad & \Cs{i,n,j+1}		\quad & \geq \quad \Ce{i,n,j} \label{cons9}\\
\forall i\leq m				\quad & \Cs{i,1,1}		\quad & \geq \quad \tau_{i} \label{cons10}\\
\forall i\leq m,n\leq N,j\leq Q_{n}	\quad & \gamma_{i}^{j}(n) 	\quad & \geq \quad 0\label{cons11}\\
\forall n\leq N				\quad & \sum_{i=1}^{m}\sum_{j=1}^{Q}\gamma_{i}^{j}(n) \quad & = \quad 1 \label{cons12}\\
\forall i \leq m			\quad & \mathrm{makespan}	\quad & \geq \quad \Ce{i,N,Q}\label{cons13}
\end{eqnarray}
}
\caption{The complete linear program.\label{linearprogram}}
\end{figure*}

Altogether, we have a linear program to be solved over the rationals,
hence a solution in polynomial time~\cite{karmarkar}. In practice,
standard packages like Maple~\cite{Map} or GLPK~\cite{glpk} will
return the optimal solution for all reasonable problem sizes.

Note that the linear program gives the optimal solution for a
prescribed number of installments for each load. We will discuss the
problem of the number of installments in the next section.

\section{Possible extensions}
\label{sec:ext}

There are several restrictions in the model of~\cite{WongVeBa05} that
can be alleviated. First the model uses \emph{uniform machines},
meaning that the speed of a processor does not depend on the task that
it executes. It is easy to extend the linear program for unrelated
parallel machines, introducing $w_{i}^{n}$ to denote the time taken by
$P_{i}$ to process a unit load of type $n$. Also, all processors and
loads are assumed to be available from the beginning.  In our linear
program, we have introduced availability dates for processors. The
same way, we could have introduced release dates for loads.
Furthermore, instead of minimizing the makespan, we could have
targeted any other objective function which is an affine combination
of the loads completion time and of the problem characteristics, like
the average completion time, the maximum or average (weighted) flow,
etc.

The formulation of the problem does not allow any piece of the $n'$th
load to be processed before the $n$th load is completely processed, if
$n' > n$. We can easily extend our solution to allow for $N$ rounds of
the $N$ loads, each load being still divided into several
installments. This would allow to interleave the processing of the
different loads.

The divisible load model is linear, which causes major problems for
multi-installment approaches. Indeed, once we have a way to find an
optimal solution when the number of installments per load is given,
the question is: what is the optimal number of installments? Under a
linear model for communications and computations, the optimal number
of installments is infinite, as the following theorem states:
\begin{theorem}
  Let us consider, under a linear cost model for communications and
  computations, an instance of our problem with one or more load and
  at least two processors. Then, any schedule using a finite number of
  installments is suboptimal for makespan minimization.
\end{theorem}

This theorem is proved by building, from any schedule, another
schedule with a strictly smaller makespan. The proof is available in
Appendix~\ref{sec:proofthm1}.

An infinite number of installments obviously does not define a
feasible solution. Moreover, in practice, when the number of
installments becomes too large, the model is inaccurate, as acknowledged
in~\cite[p. 224 and 276]{robertazzi96}. Any
communication incurs a startup cost $K$, which we express in bytes.
Consider the $n$th load, whose communication volume is $V_{comm}(n)$:
it is split into $Q_{n}$ installments, and each installment requires
$m-1$ communications.  The ratio between the actual and estimated
communication costs is roughly equal to
$\rho=\frac{(m-1)Q_{n}K+V_{comm}(n)}{V_{comm}(n)}>1$. Since $K$, $m$,
and $V_{comm}$ are known values, we can choose $Q_{n}$ such that
$\rho$ is kept relatively small, and so such that the model remains
valid for the target application. Another, and more accurate solution,
would be to introduce latencies in the model, as in~\cite{j89}. This latter
article shows how to design asymptotically optimal multi-installment
strategies for star networks. A similar approach should be used for
linear networks.

\section{Conclusion}
\label{sec:conclusion}

We have shown that a linear programming approach allows to solve all
instances of the scheduling problem addressed
in~\cite{WongVe04,WongVeBa05}. In contrast, the original approach was
providing a solution only for particular problem instances. Moreover,
the linear programming approach returns an optimal solution for any
number of installments, while the original approach was empirically
limited to very special strategies, and was often sub-optimal.

Intuitively, the solution of~\cite{WongVeBa05} is worse than the
schedule of Section~\ref{sec.ours} because it aims at locally
optimizing the makespan for the first load, and then optimizing the
makespan for the second one, and so on, instead of directly searching
for a global optimum. We did not find beautiful closed-form
expressions defining optimal solutions but, through the power of
linear programming, we were able to find an optimal schedule for any
instance.

\appendix
\section{Analytical computations for the illustrative example}
\label{annexe}

In this appendix, we prove the results stated in
Sections~\ref{oneinstallment} and~\ref{multiinstallment}.  In order to
simplify equations, we write $\alpha$ instead of $\fr{2}{1}{1}$ (i.e.,
$\alpha$ is the fraction of the first load sent from the first
processor to the second one), and $\beta$ instead of $\fr{2}{2}{1}$
(similarly, $\beta$ is the fraction of the second load sent to the
second processor).

In this research note we used simpler notations than the ones used
in~\cite{WongVeBa05}. However, as we want to explicit the solutions
proposed by~\cite{WongVeBa05} for our example, we need to use the
original notations to enable the reader to double-check our
statements. The necessary notations from~\cite{WongVeBa05} are
recalled in Table \ref{notationsbis}.

\begin{table*}[tbh]
\begin{center}
{\small
\begin{tabular}{|ll|}
\hline
$T^{n}_{cp}$	& Time taken by the standard processor ($w=1$) to compute the load $L_{n}$.\\
\hline
$T^{n}_{cm}$	& Time taken by the standard link ($z=1$) to communicate the load $L_{n}$.\\
\hline
$L_{n}$		& Size of the $n$th load, where $1 \leq n \leq N$.\\
\hline
$L_{k,n}$	& Portion of the load $L_{n}$ assigned to the $k$th installment for processing.\\
\hline
$\alpha_{n,i}^{(k)}$ & The fraction of the total load $L_{k,n}$ to $P_{i}$, where \\
		& $0\leq \alpha_{n,i}^{(k)} \leq 1, \qquad \forall i=1,\ldots, m \ \ \mathrm{and}\ \ \sum_{i=1}^{m}\alpha_{n,i}^{(k)}=1$.\\
\hline
$t_{k,n}$	& The time instant at which is initiated the first communication for the $k$th installment\\
                &  of load $L_n$ ($L_{k,n}$).\\
\hline
$C_{k,n}$	& The total communication time of the $k$th installment of load $L_{n}$ when $L_{k,n}=1$; \\
		& $C_{k,n}=\frac{T^{n}_{cm}}{L_{n}}\sum_{p=1}^{m-1}z_{p}\left(1-\sum_{j=1}^{p}\alpha_{n,j}^{(k)}\right).$\\

\hline
$E_{k,n}$	& The total processing time of $P_{m}$ for the $k$th installment of load $L_{n}$ when $L_{k,n}=1$;\\
		& $E_{k,n}=\alpha^{(k)}_{n,m}w_{m}T^{n}_{cp}\frac{1}{L_{n}}$.\\
\hline
$T(k,n)$	& The \emph{finish time} of the $k$th installment of load $L_{n}$; it is defined as the time instant\\
		& at which the processing of the $k$th installment of load $L_{n}$ ends.\\
\hline
$T(n)$		& The \emph{finish time} of the load $L_{n}$; it is defined as the time instant\\
		& at which the processing of the $n$th load ends, i.e., $T(n) = T(Q_{n})$\\
		& where $Q_{n}$ is the total number of installments required to finish processing load $L_{n}$.\\
		& $T(N)$ is the finish time of the entire set of loads resident in $P_{1}$.\\
\hline
\end{tabular}}
\caption{Summary of the notations of~~\cite{WongVeBa05} used in this paper.\label{notationsbis}}
\end{center}
\end{table*}

In the solution of~\cite{WongVeBa05}, both $P_{1}$ and $P_{2}$ have to
finish the first load at the same time, and the same holds true for
the second load.  The transmission for the first load will take
$\alpha$ time units, and the one for the second load $\beta$ time
units.  Since $P_{1}$ (respectively $P_{2}$) will process the first
load during $\lambda(1-\alpha)$ (respectively $\lambda\alpha$) time
units and the second load during $\lambda(1-\beta)$ (respectively
$\lambda\beta$) time units, we can write the following equations:
\begin{eqnarray}
\lambda(1-\alpha) & = & \alpha+\lambda\alpha \label{euxalpha}\\[2mm]
\lambda(1-\alpha)+\lambda(1-\beta) & = & (\alpha+\mathrm{max}(\beta, \lambda\alpha))+\lambda\beta\notag
\end{eqnarray}
There are two cases to discuss:
\begin{enumerate}
\item$\mathbf{\mathrm{max}(\beta, \lambda\alpha) = \lambda\alpha}$.
  We are in the one-installment case when $L_{2}C_{1,2}\leq
  T(1)-t_{1,2}$, i.e., $\beta \leq \lambda(1-\alpha)-\alpha$ (equation
  (5) in~\cite{WongVeBa05}, where $L_{2} = 1$, $C_{1,2} = \beta$,
  $T(1) = \lambda(1-\alpha)$ and $t_{1,2}=\alpha$).  The values of
  $\alpha$ and $\beta$ are given by:
  $$\alpha=\frac{\lambda}{2\lambda+1} \qquad \mathrm{and} \qquad \beta=\frac{1}{2}$$
  This case is true for $\lambda\alpha \geq \beta$, i.e.,
  $\frac{\lambda^{2}}{2\lambda+1}\geq \frac{1}{2}$ $\Leftrightarrow
  \lambda\geq \frac{1+\sqrt{3}}{2} \approx 1.366$.

  In this case, the makespan is equal to:
  $$\mathrm{makespan}_{2} = \lambda(1-\alpha)+\lambda(1-\beta) = \frac{\lambda(4\lambda+3)}{2(2\lambda+1)}.$$

  Comparing both makespans, we have:
  \begin{equation*}
    \mathrm{makespan}_{2}-\mathrm{makespan}_{1} =
    \frac{\lambda\left(2\lambda^{2}-2\lambda-1\right)}{8\lambda^{3}+12\lambda^{2}+8\lambda+2}.
  \end{equation*}

  For all $\lambda \geq \frac{\sqrt{3}+1}{2} \approx 1.366$, our
  solution is better than their one, since:
  $$\frac{1}{4}\geq \mathrm{makespan}_{2} - \mathrm{makespan}_{1} \geq 0$$
  Furthermore, the solution of~\cite{WongVeBa05} is strictly
  suboptimal for any $\lambda > \frac{\sqrt{3}+1}{2}$.

\item $\mathrm{max}(\beta, \lambda\alpha) = \beta$.  In this case,
  $P_{1}$ does not have enough time to completely send the second load
  to $P_{2}$ before the end of the computation of the first load on
  both processors.  The way to proceed in~\cite{WongVeBa05} is to send
  the second load using a multi-installment strategy.

  By using \ref{euxalpha}, we can compute the value of $\alpha$:
  $$\alpha=\frac{\lambda}{2\lambda+1}.$$
  Then we have $T(1) = (1-\alpha)\lambda =
  \frac{\lambda+1}{2\lambda+1}\lambda$ and $t_{1,2} = \alpha =
  \frac{\lambda}{2\lambda+1}$, i.e., the communication for the second
  request begins as soon as possible.

  We know from equation (1) of~\cite{WongVeBa05} that
  $\alpha^{k}_{2,1}=\alpha^{k}_{2,2}$, and by definition of the
  $\alpha$'s, $\alpha^{k}_{2,1}+\alpha^{k}_{2,2}=1$, so we have
  $\alpha^{k}_{2,i}=\frac{1}{2}$.  We also have $C_{1,2} =
  1-\alpha^{k}_{2,1} = \frac{1}{2}$, $E_{1,2} = \frac{\lambda}{2}$,
  $Y^{(1)}_{1,2}=0$, $X^{(1)}_{1,2}=\frac{1}{2}$, $H = H(1) =
  \frac{X^{(1)}_{1,2}C_{1,2}}{C_{1,2}} = \frac{1}{2}$,
  $B=C_{1,2}+E_{1,2}-H = \frac{\lambda}{2}$.

  We will denote by $\beta_{1}, \ldots, \beta_{n}$ the sizes of the
  different installments processed on each processor (then we have
  $L_{k,2} = 2\beta_{k}$).

  Since the second processor is not left idle, and since the size of
  the first installment is such that the communication ends when
  $P_{2}$ completes the computation of the first load, we have
  $\beta_{1} = T(1) - t_{1,2} = \lambda\alpha$ (see equation (27)
  in~\cite{WongVeBa05}, in which we have $C_{1,2} = \frac{1}{2}$).

  By the same way, we have $\beta_{2} = \lambda\beta_{1}$, $\beta_{3}
  = \lambda\beta_{2}$, and so on (see equation (38)
  in~\cite{WongVeBa05}, we recall that $B=\frac{\lambda}{2}$, and
  $C_{1,2}=\frac{1}{2}$):
$$\beta_{k}=\lambda^{k}\alpha$$

Each processor computes the same fraction of the second load. If we
have $Q$ installments, the total processed portion of the second load
is upper bounded as follows:
\begin{equation*}
    \sum_{k=1}^{Q}\left(2\beta_{k}\right) \leq 2\sum_{k=1}^{Q}\left(\alpha\lambda^{k}\right)
    = 2\frac{\lambda}{2\lambda+1} \lambda\frac{\lambda^{Q}-1}{\lambda-1}
    =\frac{2\left(\lambda^{Q}-1\right)\lambda^{2}}{2\lambda^{2}-\lambda-1}
\end{equation*}
if $\lambda \neq 1$, and $Q=2$ otherwise.
$$
\sum_{k=1}^{Q}\left(2\beta_{k}\right) \leq
\frac{2\lambda^{2}Q}{2\lambda+1}.
$$

We have four sub-cases to discuss:
\begin{enumerate}
\item{$0 < \lambda < \frac{\sqrt{17}+1}{8} \approx 0.64$:} Since
  $\lambda < 1$, we can write for any nonnegative integer $Q$:
  $$\sum_{k=1}^{Q}\left(2\beta_{k}\right)< \sum_{k=1}^{\infty}\left(2\beta_{k}\right)=\frac{2\lambda^{2}}{(1-\lambda)(2\lambda+1)}$$
  We have $\frac{2\lambda^{2}}{(1-\lambda)(2\lambda+1)} < 1$ for all
  $\lambda < \frac{\sqrt{17}+1}{8}$.  So, even in the case of an
  infinite number of installments, the second load will not be
  completely processed.  In other words, no solution is found
  in~\cite{WongVeBa05} for this case.

\item{$\lambda = \frac{\sqrt{17}+1}{8}$:} We have
  $\frac{2\lambda^{2}}{(1-\lambda)(2\lambda+1)} = 1$, so an infinite
  number of installments is required to completely process the second
  load.  Again, this solution is obviously not feasible.

\item{$\frac{\sqrt{17}+1}{8} < \lambda < \frac{\sqrt{3}+1}{2}$ and $\lambda \neq 1$:} In
  this case, the solution of~\cite{WongVeBa05} is better than any
  solution using a single installment per load, but it may require a
  very large number of installments.

  Now, let us compute the number of installments.  We know that the
  $i$th installment is equal to $\beta_{i}=\lambda^{i}\fr{2}{1}{1}$,
  excepting the last one, which can be smaller than
  $\lambda^{Q}\fr{2}{1}{1}$.  So, instead of writing
  $\sum_{i=1}^{Q}2\beta_{i} =
  \left(\sum_{i}^{Q-1}2\lambda^{i}\fr{2}{1}{1}\right)+2\beta_{Q} = 1$,
  we write:
  \begin{equation*}
    \sum_{i=1}^{Q}2\lambda^{i}\fr{2}{1}{1} \geq 1 \Leftrightarrow \frac{2\lambda^{2}\left(\lambda^{Q}-1\right)}{(\lambda-1)(2\lambda+1)} \geq 1
    \Leftrightarrow  \frac{2\lambda^{Q+2}}{(\lambda-1)(2\lambda+1)} \geq \frac{2\lambda^2}{(\lambda-1)(2\lambda+1)} + 1.
  \end{equation*}
  If $\lambda$ is strictly smaller than 1, we obtain:
  \begin{eqnarray*}
                    & \frac{2\lambda^{Q+2}}{(\lambda-1)(2\lambda+1)} \geq \frac{2\lambda^2}{(\lambda-1)(2\lambda+1)} + 1
    \quad \Leftrightarrow \quad 2\lambda^{Q+2} \leq 4 \lambda^{2}-\lambda-1 \\
    \Leftrightarrow & \ln(\lambda^{Q})  \leq \ln\left(\frac{4\lambda^{2}-\lambda-1}{2\lambda^{2}}\right) 
    \quad \Leftrightarrow  \quad Q\ln(\lambda)  \leq \ln\left(\frac{4\lambda^{2}-\lambda-1}{2\lambda^{2}}\right) \\
    \Leftrightarrow & Q \geq \frac{\ln\left(\frac{4\lambda^{2}-\lambda-1}{2\lambda^{2}}\right)}{\ln(\lambda)}
  \end{eqnarray*}
  We thus obtain:
  $$Q = \left\lceil \frac{\ln\left(\frac{4\lambda^{2}-\lambda-1}{2\lambda^{2}}\right)}{\ln(\lambda)} \right\rceil.$$
  When $\lambda$ is strictly greater than 1 we obtain the exact same
  result (then $\lambda-1$ and $\ln(\lambda)$ are both positive).

\item $\lambda=1$. In this case,
  $$\sum_{i=1}^{Q}2\lambda^{i}\fr{2}{1}{1} \geq 1$$
  simply leads to $Q = 2$.

\end{enumerate}
\end{enumerate}

\section{Proof of Theorem 1}
\label{sec:proofthm1}

\begin{proof}
  We first remark that in any optimal solution to our problem all
  processors work and complete their share simultaneously. To prove
  this statement, we consider a schedule where one processor completes
  its share strictly before the makespan (this processor may not be
  doing any work at all). Then, under this schedule there exists two
  neighbor processors, $P_i$ and $P_{i+1}$, such that one finishes at
  the makespan, denoted $\mathcal{M}$, and one strictly earlier. We
  have two cases to consider:
  \begin{enumerate}
  \item There exists a processor $P_i$ which finishes strictly before
    the makespan $\mathcal{M}$ and such that the processor $P_{i+1}$
    completes its share 
    exactly at time $\mathcal{M}$. 
    $P_{i+1}$
    receives all the data it processes from $P_i$. We consider any
    installment $j$ of any load $L_{n}$ that is effectively processed
    by $P_{i+1}$ (that is, $P_{i+1}$ processes a non null portion of
    the $j$th installment of load $L_{n}$). We modify the schedule as
    follows: $P_i$ enlarges by an amount $\epsilon$, and $P_{i+1}$
    decreases by an amount $\epsilon$, the portion of the $j$th
    installment of the load $L_{n}$ it processes.  Then, the
    completion time of $P_i$ is increased, and that of $P_{i+1}$ is
    decreased, by an amount proportional to $\epsilon$ as our cost
    model is linear. If $\epsilon$ is small enough, both processors
    complete their work strictly before $\mathcal{M}$. With our
    modification of the schedule, the size of a single communication
    was modified, and this size was decreased.  Therefore, this
    modification did not enlarge the completion time of any processor
    except $P_i$.  Therefore, the number of processors whose
    completion time is equal to $\mathcal{M}$ is decreased by at least
    one by our schedule modification.

  \item No processor which completes it share strictly before time
    $\mathcal{M}$ is followed by a processor finishing at time
    $\mathcal{M}$. Therefore, there exists an index $i$ such that the
    processors $P_1$ through $P_i$ all complete their share exactly at
    $\mathcal{M}$, and the processors $P_{i+1}$ through $P_m$ complete
    their share strictly earlier. Then, let the last data to be
    effectively processed by $P_i$ be a portion of the $j$th
    installment of the load $L_{n}$. Then $P_i$ decreases by a size
    $\epsilon$, and $P_{i+1}$ increases by a size $\epsilon$, the
    portion of the $j$th installment of load $L_n$ that it processes.
    Then the completion time of $P_i$ is decreased by an amount
    proportional to $\epsilon$ and the completion time of the
    processors $P_{i+1}$ through $P_m$ is increased by an amount
    proportional to $\epsilon$. Therefore, if $\epsilon$ is small
    enough, the processors $P_i$ through $P_m$ complete their work
    strictly before $\mathcal{M}$.
  \end{enumerate}
  In both cases, after we modified the schedule, there is at least one
  more processor which completes its work strictly before time
  $\mathcal{M}$, and no processor is completing its share after that
  time. If no processor is any longer completing its share at time
  $\mathcal{M}$, we have obtained a schedule with a better makespan.
  Otherwise, we just iterate our process. As the number of processors
  is finite, we will eventually end up with a schedule whose makespan
  is strictly smaller than $\mathcal{M}$. Hence, in an optimal
  schedule all processors complete their work simultaneously (and thus
  all processors work).

  We now prove the theorem itself by contradiction. Let $\mathcal{S}$
  be any optimal schedule using a finite number of installments. As
  processors $P_2$ through $P_m$ initially hold no data, they stay
  temporarily idle during the schedule execution, waiting to receive
  some data to be able to process them. Let us consider processor
  $P_2$. As the idleness of $P_2$ is only temporary (all processors
  are working in an optimal solution), this processor is only idle
  because it is lacking data to process and it is waiting for some.
  Therefore, the last moment at which $P_2$ stays temporarily idle
  under $\mathcal{S}$ is the moment it finished to receive some data,
  namely the $j$th installment of load $L_{n}$ sent to him by
  processor $P_1$.

  As previously, $Q_k$ is the number of installments of the load $L_k$
  under $\mathcal{S}$. Then from the schedule $\mathcal{S}$ we build a
  schedule $\mathcal{S}'$ by dividing in two identical halves the
  $j$th installment of load $L_{n}$. Formally:
  \begin{itemize}
  \item All loads except $L_{n}$ have the exact same installments
    under $\mathcal{S}'$ than under $\mathcal{S}$.

  \item The load $L_{n}$ has $(1+Q_{n})$ installments
    under $\mathcal{S}'$, defined as follows.

  \item The first $(j-1)$ installments of $L_{n}$ under
    $\mathcal{S}'$ are identical to the first $(j-1)$ installments of
    this load under $\mathcal{S}$.

  \item The $j$th and $(j+1)$th installment of $L_{n}$ under
    $\mathcal{S}'$ are identical to the $j$th installment of $L_{n}$
    under $\mathcal{S}$, except that all sizes are halved.

  \item The last $(Q_{n}-j)$ installments of $L_{n}$ under
    $\mathcal{S}'$ are identical to the last $(Q_{n}-j)$
    installments of this load under $\mathcal{S}$.
  \end{itemize}

  We must first remark that no completion time is increased by the
  transformation from $\mathcal{S}$ to $\mathcal{S}'$. Therefore the
  makespan of $\mathcal{S}'$ is no greater than the makespan of
  $\mathcal{S}$. We denote by \Ss{1,n,j} (respectively \Se{1,n,j})
  the time at which processor $P_1$ starts (resp.  finishes) sending
  to processor $P_2$ the $j$th installment of load $L_{n}$ under
  $\mathcal{S}$. We denote by \Cs{2,n,j} (respectively \Ce{2,n,j})
  the time at which processor $P_2$ starts (resp.  finishes) computing
  the $j$th installment of load $L_{n}$ under $\mathcal{S}$. We use
  similar notations, with an added prime, for schedule $\mathcal{S}'$.
  One can then easily derive the following properties:
  \begin{equation}
    \NSs{1,n,j}=\Ss{1,n,j}.
  \end{equation}
  \begin{equation}\label{eq:n2}
    \NSs{1,n,j+1}=\NSe{1,n,j}= \frac{\Ss{1,n,j} + \Se{1,n,j}}{2}.
  \end{equation}
  \begin{equation}\label{eq:n3}
    \NSe{1,n,j+1}=\Se{1,n,j}.
  \end{equation}
  \begin{equation}
    \NCs{2,n,j}=\NSe{1,n,j}.
  \end{equation}
  \begin{equation}\label{eq:n5}
    \NCe{2,n,j}=\NSe{1,n,j} + \frac{\Ce{2,n,j} - \Cs{2,n,j}}{2}.
  \end{equation}
  \begin{equation}\label{eq:n6}
    \NCs{2,n,j+1}=\max\{\NCe{2,n,j},\NSe{1,n,j+1}\}.
  \end{equation}
  \begin{equation}\label{eq:n7}
    \NCe{2,n,j}=\NCs{2,n,j+1} + \frac{\Ce{2,n,j} - \Cs{2,n,j}}{2}.
  \end{equation}

  Using equations \ref{eq:n2}, \ref{eq:n3}, \ref{eq:n5}, \ref{eq:n6},
  and \ref{eq:n7} we then establish that:

  \begin{multline*}
  \NCe{2,n,j}= \max\left\{
  \frac{\Ss{1,n,j} + \Se{1,n,j}}{2}\right. + \Ce{2,n,j} - \Cs{2,n,j},\\
  \Se{1,n,j} + \left.\frac{\Ce{2,n,j} - \Cs{2,n,j}}{2}
  \right\}.
  \end{multline*}

  Therefore, under schedule $\mathcal{S}'$ processor $P_2$ completes
  strictly earlier than under $\mathcal{S}$ the computation of what
  was the $j$ installment of load $L_{n}$ under $\mathcal{S}$.  If
  $P_2$ is no more idle after the time \NCe{2,n,j}, then it
  completes its overall work strictly earlier under $\mathcal{S}'$
  than under $\mathcal{S}$. On the other hand, $P_1$ completes its
  work at the same time. Then, using the fact that in an optimal
  solution all processors finish simultaneously, we conclude that
  $\mathcal{S}'$ is not optimal. As we have already remarked that its
  makespan is no greater than the makespan of $\mathcal{S}$, we end up
  with the contradiction that $\mathcal{S}$ is not optimal. Therefore,
  $P_2$ must be idled at some time after the time \NCe{2,n,j}. Then
  we apply to $\mathcal{S}'$ the transformation we applied to
  $\mathcal{S}$ as many times as needed to obtain a contradiction.
  This process is bounded as the number of communications that
  processor $P_2$ receives after the time it is idled for the last
  time is strictly decreasing when we transform the schedule
  $\mathcal{S}$ into the schedule $\mathcal{S}'$.
\end{proof}

\bibliography{biblio}
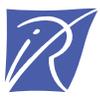
\end{document}